\documentstyle[12pt]{article}
\newcommand{\e}{{\rm e}}
\newcommand{\Xrm}{{\rm X}}
\newcommand{\Mrm}{{\rm M}}
\newcommand{\mrm}{{\rm m}}
\newcommand{\NCrm}{{\rm NC}}
\newcommand{\ug}{\; = \;}

\newcommand{\equ}{\; \equiv \;}

\newcommand{\pa}{\partial}

\newcommand{\text}{\rm}

\newcommand{\drm}{{\rm d}}

\newcommand{\Rerm}{{\rm Re}}

\newcommand{\infi}{\infty}

\newcommand{\ra}{\rightarrow}

\newcommand{\al}{\alpha}
\newcommand{\be}{\beta}
\newcommand{\ze}{\zeta}

\newcommand{\bb}{\begin{equation}}
\newcommand{\ee}{\end{equation}}
\newcommand{\bega}{\begin{eqnarray}}
\newcommand{\ega}{\end{eqnarray}}
\newcommand{\begae}{\begin{eqnarray*}}
\newcommand{\egae}{\end{eqnarray*}}

\newcommand{\h}{\hspace*{4ex}}
\newcommand{\dis}{\displaystyle}

\newcommand{\th}{\theta}

\newcommand{\om}{\omega}

\newcommand{\ovPsi}{{\overline{\Psi}}}

\newcommand{\cent}{\centerline}
\newcommand{\vs}{\vspace*}

\begin{document}

\baselineskip 0.65cm

\begin{center}

{\large {\bf New localized Superluminal solutions to the wave equations
with finite total energies and arbitrary frequencies}$^{\: (\dag)}$}
\footnotetext{$^{\: (\dag)}$  Work partially supported by MIUR and INFN
(Italy), and by FAPESP (Brazil). \ This paper did first appear as e-print
physics/0109062 [and as preprint INFN/FM--01/02 (I.N.F.N.; Frascati, 2001)]. \
E-mail addresses for contacts: recami@mi.infn.it [ER]; giz.r@uol.com.br [MZR]}

\end{center}

\vs{5mm}

\cent{ M. Zamboni-Rached, }

\vs{0.2 cm}

\centerline{{\em DMO--FEEC, State University at Campinas,
Campinas, S.P., Brazil.}}

\vs{0.5 cm}

\cent{ Erasmo Recami }

\vs{0.2 cm}

\cent{{\em Facolt\`a di Ingegneria, Universit\`a statale di Bergamo,
Dalmine (BG), Italy;}}
\cent{{\em INFN---Sezione di Milano, Milan, Italy; \ {\rm and}}}
\cent{{\em C.C.S., State University at Campinas,
Campinas, S.P., Brazil.}}

\vs{0.2 cm}

\centerline{\rm and}

\vs{0.3 cm}

\cent{ H. E. Hern\'{a}ndez-Figueroa }

\vs{0.2 cm}

\cent{{\em DMO--FEEC, State University at Campinas,
Campinas, S.P., Brazil.}}

\vs{0.8 cm}

{\bf Abstract  \ --} \ By a generalized bidirectional decomposition
method, we obtain new Superluminal localized solutions to the wave equation
(for the electromagnetic case, in particular) which are suitable for
arbitrary frequency bands; several of them being endowed with {\em finite}
total energy.  We construct, among the others, an infinite family of
generalizations of the so-called ``X-shaped" waves. \ Results of this kind
may find application in the other fields in which an essential role is played
by a wave-equation (like acoustics, seismology, geophysics, gravitation,
elementary particle physics, etc.).\\



\newpage

{\bf 1. -- Introduction}\\

\h Since many years it has been known that localized (non-dispersive)
solutions exist to the wave equation[1], endowed with
subluminal or Superluminal[2] velocities.

\h Particular attention has been paid to the localized Superluminal solutions,
which seem to exist and propagate not only in vacuum but also in media with
boundaries[3],
like normal-sized metallic waveguides[4] and possibly optical fibers.

\h It is well known that such Superluminal Localized Solutions (SLS) have
been {\em experimentally} produced in acoustics[5], in optics[6] and recently
in microwave physics[7].

\h However, all the analytical SLSs considered till now and known to us,
with one exception[8], are superposition of Bessel beams with a frequency
spectrum starting from $\nu=0$ and suitable for low frequency regions. In
this paper we shall set forth a new class of SLSs with a spectrum beginning
at any arbitrary frequency, and therefore well suited for the
construction also of high frequency (microwave, optical,...) pulses.\\

\

\

{\bf 2. -- ``$V$-cone" variables: A generalized bidirectional expansion}\\

\h Let us start from the axially symmetric solution (Bessel beam) to the wave
equation in cylindrical co-ordinates:

\

\hfill{$
\psi (\rho,z,t) \ug J_0(k\rho) \; \dis{\e^{+ik_z z} \; \e^{-i\om t}}
$\hfill} (1)

\

with the conditions

\

\hfill{$
k^2 = \dis{\frac{\om^2}{c^2} - k_z^2} \; ; \ \ \ \ \ \ k^2 \geq 0 \ ,
$\hfill} (2)

\

where $J_0$ is the zeroth-order ordinary Bessel function, and where (as
usual) $k_z$ is the longitudinal component of the wavenumber while
$k \equiv k_{\bot}$ is the wavenumber transverse component magnitude. \
The second condition (2) excludes the non-physical solutions.

\h It is essential to stress right now that the dispersion relation (2), with
positive (but not constant, a priori) $k^2$ and real $k_z$, while enforcing
the consideration of the truly propagating waves only (with exclusion of the
evanescent ones),
does allow for both subluminal and Superluminal solutions!; the latter being
the ones of interest here for us. \ Conditions (2) correspond in the
($\om,k_z$) plane to confining ourselves to the sector shown in Fig.1; that
is, to the region delimited by the straight lines $\om = \pm ck_z$.

\

A general, axially symmetric superposition of Bessel beams (with $\Phi'$ as
spectral weight-function) will therefore be:

\

\hfill{$
\Psi(\rho,z,t) \ug \dis{\int_{0}^{\infi}\drm k\;\int_{0}^{\infi}\drm \om \;
\int_{-\om /c}^{+\om /c}\drm k_z} \ \psi(\rho,z,t) \ \delta \left(k-
\sqrt{{{\om^2} \over {c^2}} - k_z^2} \right) \ \Phi'(\om,k_z;k) \ .
$\hfill} (3)

\

Notice that it is $k\geq 0$; $\om\geq 0$ and $-\om /c \leq k_z \leq +\om /c$.
The question of the negative $k_z$ values entering expansion (3) will soon
be considered below.

\h The base functions $\psi(\rho,z,t)$ can be however rewritten as

$$ \psi(\rho,\zeta,\eta) \ug J_0(k\rho) \; \exp{i[\al\ze - \be\eta]} \ , $$

where $(\al,\be)$, which will substitute in the following for the parameters
$(\om, k_z)$, are

\

\hfill{$
\al \equ \dis{{1 \over {2V}}(\om + V k_z)} \, ; \ \ \ \ \ \be \equ
\dis{{1 \over {2V}}(\om - Vk_z)} \ ,
$\hfill} (4)

\
  
in terms of the new ``$V$-cone" variables:

\

\hfill{$
\left\{\begin{array}{clr}
\zeta \equ z-Vt  \\
\eta \equ z+Vt \end{array} \right.
$\hfill} (5)

\

\h The present procedure is a generalization of the so-called ``bidirectional
decomposition" technique[9], which was devised in the past for $V = c$.

\h The ``$V$-cone" (shown in Fig.2a) corresponds in the
($\om ,k_z$) plane to the straight-lines $\om \pm Vk_z = 0$: that is, to the
lines $\al=0$ and $\be = 0$ (cf. Fig.2b); while conditions (2) become
[let us put $c=1$ {\em whenever convenient,} throughout this paper]:

\

\hfill{$
k^2 \ug V^2(\al + \be)^2 - (\al - \be)^2 \equ (\al^2 + \be^2)(V^2-1) +
2(V^2+1)\al\be \; ; \ \ \  k^2\geq 0
$\hfill} (2')

\

{\em Inside} the allowed region shown in Fig.1, we can choose for simplicity
the sector delimited by the straight-lines $\om = \pm Vk_z$, which is shown
in Fig.2b ({\em provided that} $V > 1$).

\h Let us observe that integrating over the intervals $\al, \be \geq 0$
corresponds in eq.(3) to integrating over $k_z$ between $-\om /V$ and
$+\om /V$.  But we shall choose in eq.(3) spectral weights $\Phi'(\om,k_z;k)$,
and therefore spectral weights $\Phi(\al,\be;k)$ in eq.(3') below, such as to
either eliminate or make negligible the contribution from the negative values
of $k_z$, that is, from the backwards moving waves: thus curing from the start
the problem met by the ``bidirectional decomposition" technique in connection
with the so-called non-causal components. Therefore, our SLSs will all be
physical solutions.

\h Let us recall also that each Bessel beam is associated with an
(``axicone") angle $\th$, linked to its speed by the relations[10]:

\

\hfill{$
\tan\th = \sqrt{V^2 -1}; \ \ \ \ \sin\th = \dis{{{\sqrt{V^2 -1} \over V}}}; \ \ \ \
\cos\th = \dis{{1 \over V}} \ ,
$\hfill} (6)

\

where $V \ra 1$ when $\th \ra 0$, while $V \ra \infi$ when $\th \ra \pi / 2$.

\h Therefore, instead of eq.(3) we shall consider the (more easily
integrable) Bessel beam superposition in the new variables [with $V \geq 1$]

\

\hfill{$
\begin{array}{clr}\Psi(\rho,\ze,\eta) \ug \dis{\int_{0}^{\infi}\drm k\,\int_{0}^{\infi}
\drm \al \,\int_{0}^{\infi}\drm \be \; J_0(k\rho) \; \e^{i\al\ze} \; \e^{-i\be\eta}} \, \times\;\;\;\;
\;\;\;\\ \;\;\;\; \\ \;\;\;\;\;\;\;\;\;\;\;\;\;\;\;\;\;\;\;\;\;\;\;\;
\times \, \delta\left(k-\sqrt{(\al^2 + \be^2)(V^2-1) + 2(V^2+1)\al\be}\,
\right) \; \Phi(\al,\be;k)\end{array}
$\hfill} (3')

\

where the integrations over $\al, \ \be$ between $0$ and $\infi$ just
correspond to the dashed region of Fig.2b. \ Between the spectral weights 
$\Phi$ of eq.(3') and the previous $\Phi'$ of eq.(3) it holds the relation

$$\Phi(\al,\be;k) = 2V \; \Phi'\left( V(\al+\be),\,\al-\be;\,k \right) \ ,$$

quantity $2V$ being just a dispensable multiplicative factor.

For clarity's sake, let us comment a little more on our choice of the
integration limits which enter eq.(3'). \ From eq.(3) one has that
$0 \le \om < \infty$  and $-\om/c \le k_z \le \om/c$.  From such inequalities,
and from transformations (4) written in the form $k_z=\al-\be; \ \om=V(\al+\be)$,
one easily gets that the integration limits for $\al$ and $\be$ have to obey
the three inequalities

\

\hfill{$
0 < \al+\be < \infty 
$\hfill} (4a)

\

\hfill{$
(1+V/c)\;\al \geq (1-V/c)\;\be 
$\hfill} (4b)              
            
\

and

\

\hfill{$
(1-V/c)\;\al \leq (1+V/c)\;\be  
$\hfill} (4c)  
           
\

At this point one has to carefully distinguish the case $V<c$ from $V>c$. \
In our case ($V>c$), we can write eqs.(4b) and (4c) as

\

\hfill{$
\al \geq \dis{\frac{1-V/c}{1+V/c} \; \be}   
$\hfill} (4b')  
           
\

and

\

\hfill{$
\al \geq \dis{\frac{1+V/c}{1-V/c} \; \be}     
$\hfill} (4c')  
           
\

Let us finally suppose both $\al$ and $\be$ to be positive [$\al,\be > 0$].
Inequation (4a) is then satisfied;  while the coefficients 
$(1-V/c) / (1+V/c)$  and  $(1+V/c) / (1-V/c)$  entering eqs.(4b'),(4c')
are both negatives (since $V>c$). As a consequence, the inequalities
(4b') and (4c') result to be {\em automatically} satisfied: This means that
we can actually choose $\al > 0$  and  $\be > 0$,  as we did in eq.(3'). \ In
other words, the integration limits of our eq.(3') are {\em contained}
by those discussed in connection with eq.(3), and are therefore acceptable.
Indeed, they constitute a rather suitable choice for facilitating all the
subsequent integrations.

\h We shall now go on to constructing new Superluminal Localized Solutions
for arbitrary frequencies, several of them possessing finite total energy.\\

\

\

{\bf 3. -- Some new Superluminal Localized Solutions for arbitrary
frequencies and/or with finite total energy}\\

\

{\bf 3.1} -- {\em The classical ``X-shaped solution" and its generalizations.} \\

\h Let us start by choosing the spectrum [with $a > 0$]:

\

\hfill{$
\Phi(\al,\be) \ug \delta(\be - \be') \; \e^{-a\al} \ ,
$\hfill} (7)

\

$a>0$ and $\be'\geq 0$  being constants (related to the transverse and
longitudinal localization of the pulse).

\h In the simple case when $\be'=0$, one completely dispenses with the
``non-causal" (backwards-moving) components of the bidirectional Fourier-type
expansion (3'). For the sake of clarity, let us go back to examining Fig.2b:
The $\delta(\be)$ factor in spectrum (7) does actually imply the integrations
over $\al$ and $\be$ in eq.(3') to run along the $\al$-line only; i.e., along
the $\be = 0$ straight-line (where $\om = +V k_z$). \ In this case, even more
than in the
others, it is easy to verify that the group-velocity\footnote{Let us observe
that the group velocity of the solutions considered in this paper can
a priori be evaluated through the ordinary, simple derivation of $\om$ with
respect to the wavenumber only for the infinite total energy solutions, as
in the present case. \ However, for our SSP and SMPS solutions, below, and in
general for the finite total energy Superluminal solutions, the group-velocity
cannot be calculated through that simple relation, since in those cases it
does not even exist a one-to-one function $\om = \om(k_z)$.}
of the present solution [cf. eq.(8) below] is $\pa \om / \pa k_z = 1 / \cos
\th \equiv V > 1$. \ Let us, then, choose $\be' = 0$, and observe that for $\be
= 0$ all the solutions $\Psi(\rho,\ze,\eta)$ are actually functions only of
$\rho$ and $\ze \! = \! z-Vt$. \ [Let us also notice that in empty space such
solutions $\Psi(\rho,\, \ze \! = \! z-Vt)$ can be transversely localized only
if $V\neq c$, because if $V=c$ the function $\Psi$ has to obey the Laplace
equation on the transverse planes. Let us recall that in this paper we always
assume $V>0$].

In the present case, eq.(3') can be easily integrated over $\be$ and $k$
by having recourse to identity (6.611.1) of ref.[11], yielding

\

\hfill{$
\begin{array}{clr}\Psi_{\Xrm}(\rho,\ze) \ug \dis{\int_{0}^{\infi}\drm \al \,
J_0(\rho\al\sqrt{V^2-1}) \; \e^{-\al(a-i\ze)}} \ = \\ \; \\

\;\;\;\;\;\;\, \ug \left[(a-i\ze)^2 + \rho^2(V^2-1)\right]^{-1/2} \ , \end{array}
$\hfill} (8)

\

which is exactly the classical X-shaped solution proposed by Lu \&
Greenleaf[12] in acoustics, and later on by others[12] in electromagnetism,
once relations (6) are taken into account. See Fig.3a.

\h Many other SLSs can be easily constructed; for instance, by inserting
into the weight function (7) the extra factor $\al^m$, \ namely
$\Phi(\al,\be) = \al^m \, \delta(\be) \, \exp[-a\al]$, \ where $m$ is a
non-negative integer, while it is still
$\be'=0$.  Then an infinite family of new SLSs is
obtained (for $m \geq 0$), by using this time identity (6.621.4) of the
same ref.[11]:

\

\hfill{$
\Psi_{\Xrm,m}(\rho,\ze) \ug (-i)^{m}\dis{\frac{\drm^m}{\drm \ze^m}}
\left[(a-i\ze)^2 + \rho^2(V^2-1) \right]^{-1/2}
$\hfill} (9)

\

which generalize[13] the classical $\Xrm$-shaped solution, corresponding to
$m=0$: namely, $\Psi_{\Xrm} \equiv \Psi_{\Xrm,0}$. Notice that all the
derivatives of the latter with respect to $\ze$ lead to new SLSs, all of them 
being X-shaped.

In the particular case $m=1$, one gets the SLS

\

\hfill{$
\Psi_{\Xrm,1}(\rho,\ze) \ug \dis{\frac{-i \, (a-i\ze)}{\left[(a-i\ze)^2 +
\rho^2(V^2-1)\right]^{3/2}}}
$\hfill} (10)

\

which is the first derivative of the X-shaped wave, and is depicted 
in Fig.3b.  One can notice that, by increasing $m$, the pulse becomes
more and more localized around its vertex. All such pulses travel, however,
without deforming.

\h Solution (8) is suited for low frequencies only, since its
frequency spectrum (exponentially decreasing) starts from zero. \ One can see
this for instance by writing eq.(7) in the ($\om,k_z$) plane: by eqs.(4) one
obtains

$$\Phi(\om,k_z) = \delta\left(\frac{\om-Vk_z}{2V}-\be'\right) \;
\dis{\exp[-a \, {{\om+Vk_z} \over {2V}}]}$$

and can observe that $\be'=0$ in the delta implies $\om=Vk_z$. So that the
spectrum becomes $\Phi = \exp[-a\om / V]$, which starts from zero and has a
width given by $\Delta\om=V/a$.

\h By contrast, when the factor $\al^m$
is present, the frequency spectrum of the solutions can be ``bumped" in
correspondence with any value $\om_{\Mrm}$ of the angular frequency, provided
that $m$ is large [or $a/V$ is small]: in fact, $\om_{\Mrm}$ results to be
$\om_{\Mrm} = mV / a$. \ The spectrum, then, is shifted towards higher
frequencies (and decays only beyond the value $\om_{\Mrm}$).

\h Moreover, let us mention here that also in the spectra of the following
pulses (considered in subsections 3.2 and 3.3 below) one can insert the
$\al^m$ factor; in fact, in correspondence with the spectrum

\

\hfill{$
\Phi(\al,\be) \ug \al^m \; \Phi_0(\be) \; \e^{-a\al} \ ,
$\hfill} (7')

\

one obtains {\em as further solutions} the $m$-th order derivatives of
the basic ($m=0$) solution below considered. This is due to the circumstance
that our integrations over $\al$ (as in eq.(3')) are always Laplace-type
transformations.  We shall not write them down explicitly, however, for the
sake of conciseness.

\h Different SLSs can be obtained also by {\em modifying} (still with $\be'=0$)
the spectrum (7). Some interesting solutions are reported in Appendix A.

\h Let us now construct SLSs more suited for high frequencies (always
confining ourselves to pulses well localized not only longitudinally, but
also transversely).\\

\

{\bf 3.2} -- {\em The Superluminal ``Focus-Wave Modes"} (SFWM).\\

\h Let us go back once more to spectrum (7), but examining now the general
case with $\be'\neq 0$. After integrating over $k$ and $\be$, eq.{3') yields
[$a>0$; $\be'>0; \ V>c$]:

\

\hfill{$
\Psi(\rho,\ze,\eta)\ug \dis{\e^{-i\be'\eta} \int_{0}^{\infi} \drm \al \; J_0
\left(\rho\sqrt{V^2(\al+\be')^2 - (\al-\be')^2}\right) \; \e^{-\al(a-i\ze)}} \ .
$\hfill} (11)

\

When releasing the condition $\be'=0$ we are in need also of
backwards-moving components for the construction of our pulses, since they
enter superposition (3') and therefore eq.(11). In fact, the spectrum
$\Phi=\delta(\be-\be') \, \exp[-a\al]$ does obviously entail that $\be=\be'$
and hence, by relations (4), that $\om =Vk_z + 2V\be'$. This means (see Fig.4)
that we are now integrating along the continuous line, i.e., also over the
interval $V\be' \leq \om < 2V\be'$, or $-\be' \leq k_z < 0$,
corresponding to the ``non-causal" components.  Nevertheless, we can obtain
physical solutions when making the contribution of that interval negligible,
by choosing small values of $a\be'$: so that the exponential decay of the weight
$\Phi$ with respect to $\om$ is very slow.  Actually, one can go
from the ($\al$,$\be$) space back to the ($\omega$,$k_z$) space by
use of eqs.(4), the weight being re-written (when $\be'=\be$) as
$\Phi= \exp(-a\omega /V) \cdot \exp(-a\be')$; \ wherefrom it is 
clear that\footnote{One can easily show that the condition $a \ll 1$ should
be actually replaced with the condition $a\be' \ll 1$.  In fact (see Fig.4),
the non-causal interval is $\Delta \om_{\NCrm} = V\be'$, while the total
spectral band-width is $\Delta \om = V/a$, so that the non-physical
components bring a negligible contribution to the solution in the case of
spectrum (7), provided that
$\Delta \om_{\NCrm} / \Delta \om \ll 1$, which just means $a\be' \ll 1$.}
for $a \ll 1$ the contribution of the interval $k_z \geq 0$ (or $\om \geq
2V\be'$) overruns the $k_z<0$ contribution. Notice, incidentally, that
the corresponding solutions are associated with large frequency bandwidths
and therefore to pulses with very short extension in space and in time. Let
us mention even now that the spectral weight $\Phi=\exp[-a(\omega
-V\be')/V]$ entails the frequency band-width

$$\Delta\omega \ug {V \over a} \ ,$$

a relation that we shall find to be valid (at least approximately) for all
our solutions. We shall discuss this point in Sect.5 below.

\h An analytical expression for integral (11) can be easily found for
small positive $\be'$ values, when $\be'^2 \approx 0$. Under such a
condition we obtain, by using identity (6.616.1) of ref.[11] and calling now
$X$ the classical[12] X-shaped solution (8)

\

\hfill{$
X \ug X(\rho,\eta) \equ \left[(a-i\ze)^2 + \rho^2(V^2-1)\right]^{-1/2} \ ,
$\hfill} (12)

\

we obtain the new\footnote{Notice that another, slightly different solution
---called the FXW--- appeared however as eq.(4.4) in ref.[14]} SLSs [$a>0$; $\be'\geq 0$; $V\geq c$]:

\

\hfill{$
\Psi_{\rm SFWM}(\rho,\ze,\eta) \ug \e^{-i\be'\eta} \; X \; \dis{\exp\left[
\frac{\be'(V^2+1)}{V^2-1} \, \left( (a-i\ze)-X^{-1} \right) \right]}
$\hfill} (13)

\

which for $V \ra c^+$ reduce to the well known FWM (focus-wave mode)
solutions[15], traveling with speed $c$:

\

\hfill{$
\Psi_{\rm FWM}(\rho,\ze,\eta) \ug \dis{\frac{\e^{-i\be'\eta}}{a-i\ze} \;
\exp\left[-\frac{\be'\rho^2}{a-i\ze}\right]} \ .
$\hfill} (14)

\

Our solutions (13) are a generalization of them for $V>c$; we shall call
eqs.(13) the {\em Superluminal focus wave modes} (SFWM). See Fig.5. Such
modes travel without deforming.

\h Let us emphasize that, when setting $\be'>0$, the spectrum (7) results
to be constituted (cf. Fig.4) by angular frequencies $\omega \geq V\be'$. \
Thus, our new solutions can be used to construct high frequency pulses
(e.g., in the microwave or in the optical regions): cf. also subsect.5B
below.

\h We are going now to build up suitable superpositions of
$\Psi_{\rm SFWM}(\rho,\ze,\eta)$ in order to get {\em finite} total energy
pulses, in analogy with what is currently attempted[16] for the $c$-speed
FWMs.\\

\

{\bf 3.3} -- {\em The Superluminal ``Splash Pulses"} (SSP).\\

\h In the case of the $c$-speed FWMs, in ref.[16] suitable superpositions
of them were proposed (the SPs and the ``MPS pulses") which possess finite
{\em total} energy (even without truncating them).

 \h Let us analogously go on from our solutions (13) to finite total energy
solutions, by integrating our SFWMs (13) over $\be' \;$:

\

\hfill{$
\Psi(\rho,\ze,\eta) \equ \dis{\int_0^\infi \drm \be' \; B(\be') \;
\e^{-i\be'\eta} \; X \; \exp\left[\frac{\be'(V^2+1)}{V^2-1} \, \left(
(a-i\ze)-X^{-1} \right) \right]} \ .
$\hfill} (15)

\

where it must be still $a \ll 1$, while the weight-functions $B(\be')$ must
be bumped in correspondence with {\em small} positive values of $\be'$ since
eq.(13) was obtained under the condition $\be'^2 \approx 0$. \ In the following,
for simplicity, we shall call $\be$, instead of $\be'$, the integration
variable.

\h First of all, let us choose in eq.(15) the simple weight-function [$\be'
\equiv \be$]:

\

\hfill{$
B(\be) \ = \ \e^{-b\be}
$\hfill} (16)

\

with $b \gg 0$ for the above-named reasons. Let us recall that such weight (16)
is the one yielding in the $V \ra c^+$ case the ordinary ($c$-speed) Splash
Pulses[16]; and notice that this choice is equivalent to inserting into eq.(3')
the spectral weight

\

\hfill{$
\Phi(\al,\be;k) \ \equiv \ \e^{-a\al} \; \e^{-b\be} \ .
$\hfill} (7' ')

\

\h Our {\em Superluminal Splash Pulses} (SSP) will therefore be:

\

\hfill{$
\Psi_{\rm SSP}(\rho,\ze,\eta) \ug X \; \dis{\int_0^\infi \drm \be \;
\e^{-\be (b+i\eta)} \; \e^{\be Y} \ug {X \over {b+i\eta-Y}}} \ ,
$\hfill} (17)

\

with

\

\hfill{$
Y \ \equiv \ \dis{\frac{V^2+1}{V^2-1} \; \left((a-i\ze)-X^{-1} \right)} \ .
$\hfill}

\

\h Let us repeat that our SSPs have {\em finite} total energy, as one can
easily verify; we shall come back to this result also from a geometric
point of view. \ They however get deformed while traveling, and their
amplitude decreases with time: see Figs.6a and 6b. \ It is worth mentioning
that, due to the form (7'') of the SSP spectrum, our solution (17) can be
regarded as {\em the finite energy version of the classical X-shaped
solution}.\\

\

\

{\bf 3.4} -- {\em The Superluminal ``Modified Power Spectrum"} (SMPS) {\em
pulses}.\\

\h In connection with eq.(15), let us now go on to a more general choice
for the weight-function:

\

\hfill{$
\left\{\begin{array}{clr}
&B(\be) \ug \e^{-b(\be - \be_0)} \ \ \ \ \ \ \ \ \ \ & {\rm for} \
\be \geq \be_0  \\

&B(\be) \ug 0 \ \ \ \ \ \ \ \ \ \ \ \ \ \ \ \ \ \ & {\rm for} \
0 \leq \be < \be_0
\end{array} \right.
$\hfill} (16')

\

which for $V \ra c^+$ yields the ordinary ($c$-speed) ``Modified Power
Spectrum" (MPS) pulses[16]. \ Such a choice is now equivalent to inserting
into eq.(3') for $\be \geq \be_0$ the spectrum 

\

\hfill{$
\Phi \ug \e^{-a\al} \; \e^{-b(\be - \be_0)} \ \ \ \ \ \ \ \ \ \
{\rm for} \ \be \geq \be_0 \ .
$\hfill} (7' ' ')

\

We then obtain the {\em Superluminal Modified Power Spectrum ({\rm SMPS})
pulses} as follows [for $\be_0 \ll \ 1$]:

\

\hfill{$
\Psi_{\rm SMPS}(\rho,\ze,\eta) \ug \dis{\e^{b\be_0} \; X \; \int_{\be_0}^\infi
\drm \be \; \e^{-(b+i\eta-Y)\be} \ug X \; \frac{\exp[(Y-i\eta)\be_0]}
{b-(Y-i\eta)}}
$\hfill} (18)

\

in which the integration over $\be$ runs {\em now} from $\be_0$ (no longer
from zero) to infinity.

\h It is worthwhile to emphasize that our solutions (18), like solutions (17),
possess a {\em finite total energy}.\footnote{One should recall that the first
finite energy solution, the MFXW, different from but analogous to our one,
appeared as eq.(4.6) in ref.[14].}  Even if this is easily verified, let us
address the question from an illuminating geometric point of view. \ Let us
add that their amplitude too (as for the SSPs) decreases with time: see
Figs.7a and 7b.

\

\h With reference to Fig.8, let us observe that the {\em infinite} total
energy solutions $X$, in eq.(12), and SFWM, in eq.(13), correspond to
integrations along the $\be=0$ axis (i.e., the $\al$-axis) and the
$\be=\be_0$ straight-line, respectively; that is to say, correspond to a delta
factor, $\delta(\be-\be_0)$, in the spectrum (7), where $\be' \equiv \be_0$.

\h In order to go on to the {\em finite} total energy solutions (SMPS),
eq.(18), we replaced the delta factor with the function (16'), which is zero
in the region above the $\be=\be_0$ line, while it decays[17] in the region
below (as well as along) such a line. \ The same procedure was followed by
us for the solutions SSP, eq.(17), which correspond to the particular case
$\be_0 = 0$. \ The faster the spectrum decay takes place in the region below
the $\be=\be_0$ line [i.e. $b \gg 1$], the larger the field depth\footnote{The
``depth of field" is the distance along which the pulse (approximately)
keeps its shape, besides its group-velocity; cf. refs.[16,2].}
of the corresponding pulse results to be: as we shall see in Sect.4.2C. \
Let us add that, since $b \gg 1$,  even in the present 
case the non-causal components contribution becomes negligible provided that
one chooses $a\be_0 \ll 1$; in analogy with what we obtained in the previous
SFWM case.

\ It seems important to stress also that, while the $X$ and SSP solutions,
eqs.(12) and (18), mainly consist in low-frequency (Bessel) beams, on the
contrary our solutions SFWM and SMPS, eqs.(13) and (18), can be constituted
by higher frequency beams (corresponding, namely, to $\om \geq V\be_0$). This
property can be exploited for constructing SLSs in the microwave or optics
fields, by suitable choices of the $V$ and $\be_0$ values.\\

\

\

{\bf 4. -- Geometric description of the new pulses in the} ($\om, k_z$)
{\bf plane}\\

\

{\bf 4.1} -- {\em A preliminary analysis of the localized pulses.}\\

\h Let us add some intuitive considerations about the {\em localized} solutions
$\Psi$ to the wave equation, which by our definition[18] must possess the
property

\

\hfill{$
\Psi(x,y,z;t) \ug \dis{\Psi(x,y,\,z+\Delta z_0;\,t+{{\Delta z_0} \over v})}
$\hfill} (19)

\

$v$ being the pulse propagation speed, that here can assume a priori any[1,2]
value: \ $0 \leq v < \infi$. \ Such a definition entails that the pulse
``oscillates" while propagating, it being required that it resumes
(periodically) its shape only after each space interval $\Delta z_0$, that is,
with the time interval $\Delta t_0 = \Delta z_0 / v$ (cf. refs.[18,19]).

\h Let us write the Fourier-expansion of $\Psi$ 

\

\hfill{$
\Psi(x,y,z;t) \ug \dis{\int_{-\infi}^{\infi}\drm \om \int_{-\infi}^{\infi}
\drm k_z \; \ovPsi(x,y,k_z;\om) \; \e^{i k_z z} \; \e^{-i \om t}} \ ,
$\hfill} (19a)

\

functions $\ovPsi(x,y,k_z;\om)$ and $\ovPsi(x,y,k_z;\om) \;
\exp[i(k_z \, \Delta z_0 - \om \, \Delta z_0 / v)]$ being the Fourier
transforms (with respect to the variables $z,t$) of the l.h.s. and r.h.s.
functions in eq.(19), respectively; \ where we used the translation property

$${\cal T}[f(x+a)] \ug \e^{ika} \; {\cal T}[f(x)]$$

of the Fourier transformations. \ From condition (19), we then get[18] the
fundamental constraint

\

\hfill{$
\om \ug v k_z \pm 2 n \pi \, \dis{{\frac{v}{\Delta z_0}}}
$\hfill} (20)

\

linking $\om$ with $k_z$. \ Let us explicitly mention that constraint (20)
does {\em not} imply any breakdown of the wave-equation validity.  In fact,
when inserting expression (19a) into the wave equation, one gets ---in
cylindrical plane coordinates ($\rho,\phi$)--- the physical base-solution

\

\hfill{$
\Psi(\rho,\phi,k_z;\om) \ug J_\mu(k\rho) \, \exp[i\mu\phi]
$\hfill} (19b)

\

with $\mu$ an integer and

\

\hfill{$
k^2 \ug \om^2 - k_z^2 \geq 0 \ .
$\hfill} (19c)

\

One can realize that constraint (19c), which followed from the wave equation,
is compatible with constraint (20).

\h Relation (20) is important, since it clarifies the ``spectral origin"
of the various localized solutions introduced in the past literature
(e.g., for $v=c$), which originated from superpositions performed either by
running ``along" the straight-lines (20) themselves, or in terms of spectral
weights
favouring $\om, k_z$ values not far from lines (20). In particular, in our
case, in which $v \equiv V > c$, relation (20) brings in a formal further
support of our procedures, as stated in Figs.2, 4 and 8. One may also notice
that, when the pulse spectrum does strictly obey eq.(20), the pulse depth of
field is {\em infinite} (for instance, the classical X-shaped wave and the
SFWM can be regarded as corresponding to eq.(20) with $n=0$ and $n=1$,
respectively.\footnote{On a more rigorous ground, the classical X-shaped
solution does actually correspond to eq.(20) with $\Delta z_0 \ra \infi$.
For such a reason, it does not oscillate while propagating, and travels
rigidly.  Analogously, the SSPs will not oscillate: cf. subsect.4.2.} \
While, when the spectrum is {\em only} (well) localized in the $(\om,k_z)$
plane, {\em near} one of the lines (20), the corresponding pulse has a
{\em finite} field depth (as it is the case for our SSP and SMPS solutions).
The more ``localized" the pulse spectrum is, in the $(\om,k_z)$ plane, in the
vicinity of a line (20), the longer the pulse field depth will be. \ We
shall investigate all these points more in detail, in the next subsection.\\

\

{\bf 4.2} -- {\em Spectral analysis of the new pulses.}\\

\h Let us first recall that throughout this paper it is $\om \geq 0$, and
that, whenever we deal with Superluminal or luminal speeds $V \geq c$, we
are confining ourselves (cf. Fig.2b) to the region   

\

\hfill{$
\dis{-{\om \over V} \leq k_z \leq {\om \over V}} \; ; \ \ \ \ \ \ \ \
[\om \geq 0] \ .
$\hfill} (21)

\

We are going now to generalize, among the others, what performed in ref.[18]
for the $V = c$.\\

\

\h {\bf A)} {\em Generalized $\Xrm$-shaped waves} --- In the case of the
classical X-shaped wave, {\em the spectrum} $\Phi(\al,\be) = \delta(\be) \;
\exp[-a\al]$ corresponds, because of eqs.(4), to $\Phi(\om,k_z) =
\delta(\om-Vk_z) \cdot \exp[-a(\om+Vk_z) / (2V)]$, which imposes the linear
constraint

\

\hfill{$
\om \ug V k_z \ ;
$\hfill} (20a)

\

starts from $\om = 0$; possesses the (frequency) width

$$\Delta \om \ug \dis{{\frac{V}{a}}} \ , \ \ \ \ \ \ $$

and results to be bumped for low frequencies.

\h Notice that this spectrum does exactly lies {\em along} one of the
straight-lines in Fig.4. Actually, eq.(20a) agrees with eq.(20) for
$\Delta z_0 \ra \infi$, in accord with the known fact that the pulse moves
rigidly.

\h In the case of the {\em generalized} X-pulses, while the straight-line
(20a) remains unchanged and the pulse go on being non-oscillating, the
spectrum bump moves towards higher frequencies with increasing $m$ or/and
$V/a$ (cf. subsect.3.1).\\

\

\h {\bf B)} {\em Superluminal Focus Wave Modes} --- In the case of the SFWMs,
the spectrum $\Phi(\al,\be) = \delta(\be-\be') \; \exp[-a\al]$ corresponds
(because of eqs.(4)) to $\Phi(\om,k_z) = \delta(\om-Vk_z-2V\be')) \cdot
\exp[-a(\om+Vk_z) / (2V)]$, which imposes the linear constraint

\

\hfill{$
\om \ug V k_z + 2V\be' \ .
$\hfill} (20b)

\

The minimum value of $\om$ is given (see Fig.4 and relation (21)) by the
intersection of the straight-lines (20b) and $\om=-Vk_z$.  This spectrum
starts from $\om_{\rm min} = V \be'$ and possesses the (frequency)
width

$$\Delta \om \ug \dis{{V \over a}} \ .$$

\h Notice that, once more, the spectrum runs exactly along the line (20b).
By comparing eq.(20b) with eq.(20), one gets that for these oscillating
solutions the {\em periodicity} space and time intervals are

$$\Delta z_0 \ug \dis{{\pi \over \be'}} \ ; \ \ \ \ \ \Delta t_0 \ug
\dis{{\pi \over {V\be'}}} \ .$$

Let us recall from subsect.3.2 and Fig.4 that it must be $a\be' \ll 1$ in
order to make negligible the non-causal component contribution (in the
two-dimensional expansion). As mentioned in subsect.3.2, the
relation $\om \geq V\be'$ can be exploited for obtaining high frequency
SLSs.\\

\

\h {\bf C)} {\em Superluminal Splash Pulses} --- In the case of the SSPs, the
spectrum $\Phi(\al,\be) = \exp[-b\be] \; \exp[-a\al]$ corresponds (because of
eqs.(4)) to $\Phi(\om,k_z) = \exp[-b(\om-Vk_z) / (2V)] \cdot \exp[-a(\om+Vk_z)
/ (2V)]$. \ This time the spectrum is {\em no longer} exactly localized over
one of the lines (20); however, if we choose $b \gg 1$ and $a \ll 1$, such a
choice together with condition (21) implies $\Phi(\om,k_z)$ to be well
localized in the neighborhood of the line

\

\hfill{$
\om \ug V k_z \ ,
$\hfill} (20c)

\

besides being almost exclusively composed of causal components. All this can
be directly inferred also from the form of $\Phi(\al,\be)$, in connection
with Fig.8. The spectrum starts from $\om_{\rm min} = 0$, with the frequency
width

$$\Delta \om \simeq \dis{{V \over a}} \ . \ \ \ \ \ \ $$

\h  Equation (20) can be compared with eq.(20c) only when $b \gg 1$; \ under
such a condition, we obtain that $\Delta z_0 \ra \infi$. \ However, since
$b$ can be large but not infinite, the pulse is expected to be endowed in
reality with a slowly decaying amplitude, as shown below in subsect.5.2.\\

\

\h {\bf D)} {\em Superluminal Modified Power Spectrum Pulses} --- In the case
of the SMPS pulses, the spectrum is $\Phi(\al,\be) = 0$ for $0 \leq \be
< \be_0$, and $\Phi(\al,\be) =\exp[b(\be-\be_0)] \; \exp[-a\al]$ for $\be
\geq \be_0$. \ Under the condition $b \gg 1$ it is $\be \simeq \be_0$, that
is to say, the spectrum is well localized (as it follows from eqs.(4)) in the
vicinity of the straight-line

\

\hfill{$
\om \ug V k_z + 2 V \be_0 \ .
$\hfill} (20d)

\

To enforce causality, we choose (as before) also $a\be_0 \ll 1$.  Like in the SFDW
pulse case, the spectrum starts from $\om_{\rm min} = V\be_0$, with the frequency
width

$$\Delta \om \simeq \dis{{V \over a}} \ .$$

Once more, in the case when $b \gg 1$, one can compare eq.(20) with eq.(20d),
obtaining $\Delta z_0 \simeq \pi / \be_0$ and $\Delta t_0 \simeq \pi /
(V\be_0)$. \ Under the condition $b \gg 1$, the pulse is expected to possess
a long depth of field, and propagate along it (in an oscillating way) with
a maximum amplitude almost constant: we shall look more in detail at this
behaviour in subsect.5.3.\\

\

\

{\bf 5. -- Some exact (Superluminal localized) solutions, and their field
depth}\\

\

\h To inquiring more in detail into the field depth of our SLSs, we can
confine ourselves to the propagation straight-line $\rho = 0$. Then, we can
find {\em exact analytic} solutions holding for any value of $\be'$, without
having to assume $\be'$ to be small, as we had on the contrary to assume for
the SFWM, the SST and the SMPS solutions (see Sect.3, subsections 1, 2, 3).
In fact, one is confronted with a simple integration of the type

\

\hfill{$
\Psi(\rho=0,\ze,\eta) \ug \dis{\int_{0}^{\infi}\drm \al \, \int_{0}^{\infi}
\drm \be \; \e^{-i\be\eta} \; \e^{i\al\ze} \; \Phi(\al,\be)} \ .
$\hfill} (3' ')

\

Let us first study the infinite total energy solutions: namely, our
SFWMs (skipping the generalized X-type solutions).\\

\

{\bf 5.1} -- {\em The case of the Superluminal Focus Wave Modes.}\\

\h In the case of the SFWMs, solution (11) may be integrated for $\rho=0$,
without imposing the small $\be_0 \equiv \be'$ 
approximation.\footnote{Also in the case of the SMPS pulses, below, we shall  
arrive at analytical solutions without any need of imposing the condition
that $\be_0 \equiv \be'$ be small.}
In fact, by choosing $\Phi$ like in eq.(7), one obtains

\

\hfill{$
\Psi_{\rm SFWM}(\rho=0,\ze,\eta) \ug \dis{\e^{-i\be_0\eta} \int_{0}^{\infi}
\drm \al \; \e^{i\al\ze} \; \e^{-a\al} \ug \e^{-i\be_0\eta} \; (a-i\ze)^{-1}}
$\hfill} (11a)

\

whose square magnitude $|\Psi|^2 = (a^2+\ze^2)^{-1}$ reveals that
$\Psi_{\rm SFWM}$ is endowed with an infinite depth of field.

\h Due to the linearity of the wave equation, both the real and the imaginary
part of  eq.(11a), as well as of all our (complex) solutions, are themselves
{\em solutions} of the wave equation. \ In the following we shall confine
ourselves to investigating the behaviour of the real part.

\h In the case of eq.(11a) it is

\

\hfill{$
\Rerm\left[\Psi_{\rm SFWM}(\rho=0,\ze,\eta)\right] \ug \dis{{{a \,
\cos(\be_0\eta) + \ze \, \sin(\be_0\eta)} \over {a^2 + \ze^2}}} \ .
$\hfill} (11b)

\

The center $C$ of such a pulse (where the pulse reaches its maximum value,
$M$, {\em oscillating} in space and time) corresponds to $z=Vt$, that is,
to $\ze=0$ and $\eta=2z$; its value being

\

\hfill{$
M_{\rm SFWM} \ug \dis{{{\cos(2\be_0z)} \over a}} \ .
$\hfill} (11c)

\

Notice that: \ (i) at $C$ one meets the maximum value $M$ of the whole
three-dimensional pulse: \ (ii) quantity $M$ is a periodic function of $z$
(and $t$), with ``wavelength" $\Delta z_0$ (and oscillation period $\Delta
t_0$) given by

\

\hfill{$
\Delta z_0 \ug \dis{{\pi \over \be_0}} \ ; \ \ \ \ \ \Delta t_0 \ug
\dis{{\pi \over {V\be_0}}} \ ,
$\hfill} (11d)

\

respectively: in agreement with what anticipated in subsect.4.2-B.

\ The delta function entering our spectrum (7), entailing that $\be=\be_0$,
requires that

\

\hfill{$
\om \ug V k_z + 2 V \be_0
$\hfill} (22)

\

which is nothing but the straight-line $\be=\be_0$ of Fig.8; \ this fact
implying by the way (as we already saw) and infinite field depth, in
accordance with the previous considerations in subsect.3.4.

\h By comparing eq.(22) with the important ``localization constraint" (20),
with $n=1$, we just obtain the value $\Delta z_0$ of eq.(11d).  In other
words, the previously got relations (11d) are exactly what needed for the
localization properties (non-dispersiveness) of our SFWMs.

\h Finally, let us  examine the longitudinal localization of our oscillating
beams. For simplicity, let us analyse the ``dispersion" of the beam when its
amplitude is maximal; let us therefore skip considering the oscillations and
go on to the pulse {\em magnitude}: one gets for the pulse half-height
full-width the value $D=2\sqrt{3}a$ in the case of the magnitude itself, and

\

\hfill{$
D \ug 2a
$\hfill} (23)

\

in the case of the {\em square} magnitude.  Let us adhere to the latter
choice in the following, due to a widespread use.\\

\

{\bf 5.2} -- {\em The finite total energy solutions.}\\

\h Let us now go on to the {\em finite} total energy solutions:\\

\

\h {\bf a)} {\em The case of the Superluminal Splash Pulses} --- In the case
of the SSPs with $\rho=0$, one has to insert into eq.(3' ') the spectrum
(7' '), namely $\Phi = \exp[-a\al] \; \exp[-b\be]$.  By integrating, we
obtain

\

\hfill{$
\Psi_{\rm SSP}(\rho=0,\ze,\eta) \ug \dis{\left[(a-i\ze)(b+i\eta)\right]^{-1}}
\ ,
$\hfill} (17a)

\

whose {\em real part} is

\

\hfill{$
\Rerm\left[\Psi_{\rm SSP}(\rho=0,\ze,\eta)\right] \ug \dis{ {{ab+\eta\ze}
\over {(ab+\eta\ze)^2 + (a\eta-b\ze)^2}} } \ .
$\hfill} (17b)

\

\h Let us explicitly observe that the chosen spectrum, by virtue of eqs.(4),
entails that these solutions (17a,b) do {\em not} oscillate, which correspond
to $\Delta z_0 \ra \infi$ and $\Delta t_0 \ra \infi$ in eqs.(20): in agreement
with what anticipated in subsect.4.2-C. \ Actually, the SSPs are the {\em finite
energy} version of the classical X-shaped pulses.

\h The maximum value $M$ of eq.(17b) (a not oscillating, but slowly decaying
only, solution) still corresponds to putting $z=Vt$, that is, to setting
$\ze=0$ and $\eta=2z$:

\

\hfill{$
M_{\rm SSP} \ug \dis{ {b \over a} \cdot {1 \over {b^2+4z^2}} } \ .
$\hfill} (17c)

\

Initially, for $z=0, \ t=0$, we have $M=(ab)^{-1}$. \ If we now {\em define
the field-depth Z as the distance over which the pulse's amplitude is 90\%
at least of its initial value}, then we obtain the depth of field

\

\hfill{$
Z_{\rm SSP} \ug \dis{{b \over 6}}
$\hfill} (24)

\

which shows the dependence of $Z$ on $b$, namely, the dependence of $Z$ on
the spectrum localization in the surroundings of the straight-line
$\om=Vk_z$: Cf. also subsect.3.3.

\h At last, the longitudinal localization will be approximately given by

\

\hfill{$
D \approx 2a \ ;
$\hfill} (25)

\

namely, it is still given (for $a \ll 1$ and $b \gg 1$) by eq.(23).  Notice
that, since solution (17a) does not oscillate, the same will be true for its
real part, eq.(17b), as well as for the square {\em magnitude} of eq.(17a): as
it can be straightforwardly verified. Of course, equation (25) holds for
$t=0$. During the pulse propagation, the longitudinal localization $D$ seems
to increase, while the amplitude $M$ decreases. Indeed, our preliminary
calculations have verified that the $D$-increase rate is approximately equal
to the $M$-decrease rate; so much so we obtain (practically) the same field
depth, eq.(24), when requesting the longitudinal localization to suffer
a limited increase (e.g., by 10\% only).\\

\

\h {\bf b)} {\em The case of the Superluminal Modified Power Spectrum
pulses} --- In the case of the SMPS pulses with $\rho=0$, one has to insert
into eq.(3' ') the spectrum (7' ' '), namely $\Phi = \e^{-a\al} \,
\e^{-b(\be - \be_0)}$, with $\be \geq \be_0$. By integration, one gets

\

\hfill{$
\Psi_{\rm SMPS}(\rho=0,\ze,\eta) \ug \dis{\e^{-i\be_0\eta} \; \left[(a-i\ze)
(b+i\eta)\right]^{-1}} \ ,
$\hfill} (18a)

\

whose real part is easily evaluated. \ These pulses do oscillate while
traveling. Their field depth, then calculated by having recourse to the pulse
square magnitude, happens still to be

\

\hfill{$
Z_{\rm SMPS} \ug \dis{{b \over 6}}
$\hfill} (26)

\

like in the SSP case. Even the longitudinal localization of the square
amplitude results approximately given, for $t=0$, by

\

\hfill{$
D \approx 2a
$\hfill} (27)

\

as in the previous cases.

\h The field depth (26) depends only on $b$. \ However, the {\em behaviour} of
the propagating pulse {\em changes} with the $\be_0$-value change, besides
with $b$'s. \ Let us examine the maximum amplitude of the real part of
eq.(18a), which for $z=Vt$ writes (when $\ze=0$ and $\eta=2z$):

\

\hfill{$
M_{\rm SFWM} \ug \dis{{1 \over {ab}} \; {{\cos(2\be_0z) + 2[z/b] \,
\sin(2\be_0 z)} \over {1+4[z/b]^2}}} \ .
$\hfill} (18b)

\

Initially, for $z=0, \ t=0$, one has $M = (ab)^{-1}$ like in the SSP case.

\h From eq.(18b) one can infer that:

\h {\bf (i)} when $z/b \ll 1$, namely, when $z < Z$, eq.(18b) becomes

\

\hfill{$
M_{\rm SMPS} \; \simeq \; \dis{{{\cos(2\be_0z)} \over {ab}}} \ , \ \ \ \ \ \ \ \
[{\rm for} \ z<Z]
$\hfill} (28)

\

and the pulse does actually oscillate {\em harmonically} with wavelength
$\Delta z_0 = \pi / \be_0$ and period $\Delta t_0 = \pi/(V\be_0)$, all along
its field depth: In agreement with what anticipated in subsect.4.2-D.

\h {\bf (ii)} when $z/b > 1$, namely, when $z > Z$, eq.(18b) becomes

\

\hfill{$
M_{\rm SMPS} \; \simeq \; \dis{{{\sin(2\be_0z)} \over {ab}} \; {1 \over {2 \,
[z/b]}}}  \ \ \ \ \ \ \ \
[{\rm for} \ z>Z]
$\hfill} (28')

\

Therefore, {\em beyond} its depth of field, the pulse go on oscillating
with the same $\Delta z_0$, but {\em its maximum amplitude decays}
proportionally to $z$ (the decay coefficient being $b/2$).\\

\h Last but not least, let us add the observation that results of this kind
may find application in the other fields in which an essential role is played
by a wave-equation (like acoustics, seismology, geophysics, relativistic
quantum physics, gravitational waves).\\

{\bf Acknowledgements}\\

The authors are grateful, for stimulating discussions and kind cooperation,
to A.Arecchi, C.E.Becchi, M.Brambilla, C.Cocca, R.Collina, R.Colombi,
G.C.Costa, P.Cotta-Ramusino, F.Fontana, G.C.Ghirardi, L.C.Kretly, L.Lugiato,
K.Z.N\'obrega, G.Pedrazzini, G.Salesi, A.Shaarawi and J.W.Swart, as well as
J.Madureira and M.T.Vasconselos. \ This paper first appeared as e-print
physics/0109062.

\newpage

{\centerline{{\bf APPENDIX A}}

\

\centerline{{\em Further families of ``X-type" Superluminal localized
solutions}}

\

\h As announced in subsect.3.1, let us mention in this Appendix that one can
obtain new SLSs by considering {\em for instance} the following modifications
(still with $\be'=0$ of the spectrum (7), with $a$, $d$ arbitrary constants:

\

\hfill{$
\Phi(\al,\be;k) \ug \delta(\be) \; J_0(2d\sqrt{\al}) \; \e^{-a\al}
$\hfill} (A.1a)

\

\hfill{$
\Phi(\al,\be;k) \ug \delta(\be) \; \sinh(\al d) \; \e^{-a\al}
$\hfill} (A.1b)

\

\hfill{$
\Phi(\al,\be;k) \ug \delta(\be) \; \cos(\al d) \; \e^{-a\al}
$\hfill} (A.1c)

\

\hfill{$
\Phi(\al,\be;k) \ug \delta(\be) \; \dis{ { {\sin\al d} \over \al }  \; \e^{-a\al} }
$\hfill} (A.1d)

\

Let us call $X$, as in eq.(8), the classical X-shaped solution

$$X \equ \left[(a-i\ze)^2 + \rho^2(V^2-1)\right]^{{1 \over 2}} \ .$$

One can obtain from those spectra the new, different Superluminal localized
solutions, respectively:

\

\hfill{$
\begin{array}{clr}\Psi(\rho,\ze) \ug X \cdot J_0(\rho d^2 \, \sqrt{V^2-1} \; X^2)
\times \\ \; \\

\;\;\;\;\;\;\, \ \ \times \exp\left[-(a-i\ze) \, d^2 \, X^2\right] \ , \end{array}
$\hfill} (A.2a)

\

got by using identity (6.6444) in ref.[11];

\

\hfill{$
\Psi(\rho,\ze) \ug \dis{ { {2d(a-i\ze) \sqrt{2(X^{-2}+d^2)}} \over {(X^{-2}+d^2) -
4 d^2 (a-i\ze)^2} } } \ ,
$\hfill} (A.2b)

\

for $a > |d|$, by using identity (6.668.1) of ref.[11];

\

\hfill{$
\Psi(\rho,\ze) \ug \left[ \dis{ { {X^{-2}-d^2+\sqrt{(X^{-2}-d^2)^2+4d^2(a-i\ze)^2}}
\over {2 [ (X^{-2}-d^2)^2+4d^2(a-i\ze)^2 ]} } } \right]^{{1 \over 2}}
$\hfill} (A.2c)

\

by using identity (6.751.3) of ref.[11]; \ and

\

\hfill{$
\begin{array}{clr}\Psi(\rho,\ze) \ug {\sin}^{-1} \ 2d \, \left[ \dis{ \sqrt{X^{-2}
+d^2+2\rho d \sqrt{V^2-1} } \; \ + } \right. \\

\ \ + \ \left. \dis{ \sqrt{X^{-2}+d^2-2\rho d \sqrt{V^2-1}} } \; \right] \ , \end{array}
$\hfill} (A.2d)

\

for $a>0$ and $d>0$, by using identity (6.752.1) of ref.[11].

\

\h Let us recall that, due to the choice $\be'=0$ and the consequent presence
of a $\delta(\be)$ factor in the weight, all such solutions are completely
physical, in the sense that they {e\ don't} get {\em any} contribution from
the non-causal components (i.e., from waves moving backwards). \ In fact,
these new solutions are functions of $\rho, \ze$ only (and not of $\eta$). \
In particular, solutions (A.2b), (A.2c), (A.2d), as well as others easily
obtainable, are functions of $\rho$ via quantity $X$ only.  This may suggest
to go on from the variables $(\rho, \ze)$ to the variables $(X, \ze)$ and
write down {em the wave equation itself} in the new variables:  Some related
results and consequences will be exploited elsewhere.

%
%
%
%
%
%
%
%
%
%
%

\newpage

\centerline{{\bf Figure captions}}

Fig.1 -- Geometrical representation, in the plane ($\om,k_z$), of our 
conditions (2), with $\om \geq 0$: see the text. It is essential to notice that the dispersion
relation (2), with positive (but not constant, a priori) $k^2$ and real
$k_z$, while enforcing the consideration of the truly propagating waves only
(with exclusion of the evanescent ones), does allow for both subluminal and
Superluminal solutions; the latter being the ones of interest for us. \
Conditions (2) correspond to confining ourselves to the sector delimited
by the straight lines $\om = \pm ck_z$.\\

Figs.2 -- The ``$V$-cone" (shown in figure $a$) corresponds in the
($\om ,k_z$) plane to the straight-lines $\om \pm Vk_z = 0$. \ Inside the
allowed region, shown in Fig.1, we choose for simplicity (see the text)
the sector depicted in figure $b$. \ We assume $V > 1$ and confine ourselves
to $\om \geq 0$.

Figs.3 -- In Fig.3a it is represented (in arbitrary units) the square
magnitude of the ``classical", $X$-shaped Superluminal Localized Solution (SLS)
to the wave equation[12], with $V=5c$ and $a=0.1 \;$m: cf. eqs.(8) and (6). \
An infinite family of SLSs however exists, which generalize the classical
$X$-shaped solution; the Fig.3b depicts the first of them (its first
derivative) with the same parameters: see the text and eq.(10). The
successsive solutions in such a
family are more and more localized around their vertex.  Quantity $\rho$ is
the distance in meters from the propagation axis $z$, while quantity $\ze$
is the ``$V$-cone" variable (still in meters) $\ze \equiv z-Vt$, with
$V \geq c$. \ Since all
these solutions depend on $z$ only ia the variable $\ze$, they propagate
``rigidly", i.e., without distortion (and are called ``localized", or
non-dispersive, for such a reason).  In this paper we assume propagation in
the vacuum (or in a homogeneous medium).\\

Fig.4 -- When releasing the condition $\be'=0$ (see the text), which
excluded the ``backwards-traveling" components, one has to integrate
in eq.(11) along the half-line $\om = Vk_z+\be'$, namely, also along the 
``non-causal" interval $V\be' < \om < 2V\be'$. \ We can obtain physical
solutions, however, by making negligible the contribution of the
unwanted interval, i.e., by choosing small values of $a$.  This can be
even more easily seen in the ($\om, k_z$) plane.\\

Fig.5 -- Representation of our Superluminal Focus Wave Modes (SFWM), eq.(13),
which are a generalization of the ordinary FWMs. The depicted pulse
corresponds to $V=5c$, $a=0.001 \;$m; $\beta' = 1/(100 \, \mrm)$, and to
arbitrary time $t$ (since these solutions too travel without deforming). \
Such solutions correspond to high frequency (microwave, optical,...)
pulses: see the text. \ The meaning of $\rho$, $\ze$, etc., is given in
the caption of Fig.3.\\

Figs.6 --  Representation of our Superluminal Splash Pulses (SSP), eq.(17). \
They are suitable superpositions of SFWMs (cf. Fig.5), so that their total
energy is {\em finite} (even without any truncation). They however get deformed
while propagating, since their amplitude decreases with time. \ In Fig.6a we
represent, for $t=0$, the pulse corresponding to $V=5c$, $a=0.001 \;$m,
and $b=200 \;$m. \ In Fig.6b it is depicted the same pulse after having
traveled 50 meters.\\

Figs.7 -- Representation of our Superluminal Modified Power Spectrum (SMPS)
pulses, eq.(18). Also these beams possess {\em finite} total energy, and
therefore get deformed while traveling. \ Fig.7a depicts the shape of the
pulse, for $t=0$, with $V=5c$, $a=0.001 \;$m, $b=100 \;$m, and $\beta_0 =
1/(100 \, \mrm)$. \ In Fig.7b it is shown the same pulse after a 50 meters
propagation.\\

Figs.8 -- From a geometric point of view, our infinite total energy SLSs,
i.e., the $X$-solutions, eq.(12), and the SFWMs, eq.(13), correspond ---see
the text--- to integrations along the $\be=0$ axis, or $\al$-axis, and the
$\be=\be'$ straight-line, respectively. \ In order to go on to the {\em finite}
total-energy SLSs, we had to replace the $\delta(\be-\be')$ factor in the
spectrum (7) with the function (16'), which is different from $0$ in the region
along and below the $\be=\be'$ line and suitably {\em decays} therein.  The
faster the spectrum decays (below the $\be=\be'$ line), the larger
the field depth of the pulse results to be. \ In such a manner we obtained
the SMPSs, eq.(18), as well as the SSPs, which just correspond to the
particular case $\be'=0$.

\newpage

REFERENCES\\

[1] See, e.g., R.Courant and D.Hilbert: {\em Methods of Mathematical Physics}
(J.Wiley; New York, 1966), vol.2, p.760; \ J.A.Stratton:  {\em
Electromagnetic Theory}  (McGraw-Hill; New York, 1941), p.356; \ H.Bateman:
{\it Electrical and Optical Wave Motion} (Cambridge Univ.Press; Cambridge,
1915). \ See also: V.K.Ignatovich: Found. Phys. 8 (1978) 565; \
J.N.Brittingham: J. Appl. Phys. 54 (1983) 1179; \ R.W.Ziolkowski: J. Math.
Phys. 26 (1985) 861; \ J.Durnin, J.J.Miceli and J.H.Eberly: Phys. Rev.
Lett. 58 (1987) 1499; Opt. Lett. 13 (1988) 79; \ A.M.Shaarawi,
I.M.Besieris and R.W.Ziolkowski: J. Math. Phys. 31 (1990) 2511; \
A.O.Barut et al.: Phys. Lett. A143 (1990) 349; Found. Phys. Lett.
3 (1990) 303; Found. Phys. 22 (1992) 1267; Phys. Lett. A180 (1993) 5; A189
(1994) 277; \ P.Hillion: Acta Applicandae Matematicae 30 (1993) 35; \
R.Donnelly and R.W.Ziolkowski: Proc. Roy. Soc. London A440 (1993) 541; \
J.Vaz and W.A.Rodrigues: Adv. Appl. Cliff. Alg. S-7 (1997) 457; \ S.Esposito:
Phys. Lett. A225 (1997) 203.\hfill\break 

[2] E.Recami: Physica A252} (1998) 586; \ J.-y.Lu, J.F.Greenleaf
and E.Recami: ``Limited diffraction solutions to Maxwell (and Schroedinger)
equations'', Lanl Archives \# physics/9610012 (Oct.1996); \ R.W.Ziolkowski,
I.M.Besieris and A.M.Shaarawi: J. Opt. Soc. Am. A10 (1993) 75; \
J.-y.Lu and J.F.Greenleaf: IEEE Trans. Ultrason. Ferroelectr. Freq. Control
39 (1992) 19. \ Cf. also E.Recami, in {\em Time's Arrows, Quantum
Measurement and Superluminal Behaviour}, ed. by D.Mugnai, A.Ranfagni and
L.S.Shulman (C.N.R.; Rome, 2001), pp.17-36.\hfill\break

[3] M.Zamboni-Rached and H.E.Hern\'andez-Figueroa: Optics Comm. 191 (2000)
49. \ From the experimental point of view, cf. S.Longhi, P.Laporta, M.Belmonte
and E.Recami: ``Measurement of superluminal optical tunnelling in double-barrier
photonic bandgaps", Phys. Rev. A65 (2002) 046610.\hfill\break

[4] M.Zamboni, E.Recami and F.Fontana: ``Localized Superluminal solutions to
Maxwell equations propagating along a normal-sized waveguide", Phys. Rev.
E64 (2001) 066603.\hfill\break

[5] J.-y.Lu and J.F.Greenleaf: IEEE Trans. Ultrason. Ferroelectr. Freq.
Control 39 (1992) 441:  In this case the beam speed is larger than the
{\em sound} speed in the considered medium.\hfill\break

[6] P.Saari and K.Reivelt: ``Evidence of X-shaped propagation-invariant
localized light waves", Phys. Rev. Lett. 79 (1997) 4135.\hfill\break

[7] D.Mugnai, A.Ranfagni and R.Ruggeri: Phys. Rev. Lett. 84 (2000)
4830. \ For a panoramic review of the ``Superluminal" experiments, see
E.Recami: [Lanl Archives physics/0101108], Found. Phys. 31 (2001)
1119.\hfill\break

[8] P.Saari and H.S\~{o}najalg: Laser Phys. 7 (1997) 32.\hfill\break

[9] A.Shaarawi, I.M.Besieris and R.W.Ziolkowski: J. Math. Phys. 30 (1989)
1254; \ A.Shaarawi, R.W.Ziolkowski and I.M.Besieris: J. Math. Phys. 36 (1995)
5565.\hfill\break

[10] E.Recami et al.: Lett. Nuovo Cim. 28 (1980) 151; 29 (1980) 241; \
A.O.Barut, G. D.Maccarrone and E.Recami: Nuovo Cimento A71 (1982) 509. \
See also E.Recami: Rivista N. Cim. 9(6) (1986) 1--178; \ E.Recami: ref.[2]; \
E.Recami, F.Fontana and R.Garavaglia: Int. J. Mod. Phys. A15 (2000) 2793; \
and E.Recami et al.: Il Nuovo Saggiatore 2(3) (1986) 20; 17(1-2) (2001) 21.
\hfill\break

[11] I.S.Gradshteyn and I.M.Ryzhik: {\em Integrals, Series and Products},
4th edition (Ac.Press; New York, 1965).\hfill\break 

[12] J.-y.Lu and J.F.Greenleaf: in refs.[2]; \ E.Recami: in refs.[2].
\hfill\break

[13] Similar solutions were considered in A T. Friberg, J. Fagerholm and
M.M.Salomaa: Opt. Commun. 136 (1997) 207; and J.Fagerholm, A.T.Friberg,
J.Huttunen, D.P.Morgan and M.M.Salomaa: Phys. Rev. E54 (1996) 4347; 
as well as in P. Saari: in {\it Time's Arrows, Quantum Measurements
and Superluminal Behavior}, ed. by D.Mugnai et al. (C.N.R.; Rome, 2001),
pp.37-48.\hfill\break

[14] I.M.Besieris, M.Abdel-Rahman, A.Shaarawi and A.Chatzipetros:
Progress in Electromagnetic Research (PIER) 19 (1998) 1.\hfill\break

[15] R.W.Ziolkowski: Phys. Rev. A39 (1989) 2005; J. Math. Phys. 26 (1985) 861; \
P.A.Belanger: J. Opt. Soc. Am. A1 (1984) 723; \ A.Sezginer: J. Appl. Phys. 57
(1985) 678.\hfill\break

[16] R.W.Ziolkowski: ref.[15]; \ A.Shaarawi, I.M.Besieris and R.W.Ziolkowski:
ref.[9]; \ I.M.Besieris, M.Abdel-Rahman, A.Shaarawi and A.Chatzipetros:
ref.[14]. \ Cf. also A.M.Shaarawi and I.M.Besieris: J. Phys. A: Math.Gen. 33
(2000) 7227; 33 (2000) 7255; 33 (2000) 8559; Phys. Rev. E62 (2000) 7415.
\hfill\break

[17] The relaxation of the spectral delta correlation has been discussed
(even if for a different set of coordinates, i.e., over a different plane)
also in the paragraphs associated with eqs.(3.5),(3.6) in A.M.Shaarawi: J.
Opt. Soc. Am. A14 (1997) 1804-1816, and with eqs.(4.2),(4.3) in A.M.Shaarawi,
I.M.Besieris, R.W.Ziolkowski and R.M.Sedky: J. Opt. Soc. Am. A12 (1995)
1954-1964; \ while the need for a relaxation of that kind in order to get
finite energy solutions was mentioned (as we already said) in ref.[14],
besides ref.[18].\hfill\break

[18] M.Zamboni-Rached: ``Localized solutions: Structure and Applications",
M.Sc. thesis (Phys. Dept., Campinas State University, 1999).\hfill\break

[19] Cf. also Ruy H.A.Farias and E.Recami: ``Introduction of a Quantum of Time
(``chronon"), and its Consequences for Quantum Mechanics", Lanl Archive \#
quant-ph/9706059, and refs. therein; \ P.Caldirola: Rivista N. Cim. 2 (1979),
issue no.13.

\end{document}